\documentclass[]{pasj01}
\usepackage{natbib}

\newcommand{\FeH}{$\mbox{[Fe/H]}$} 
\newcommand{\AB}[2]{$\mbox{[#1/#2]}$} 

\newcommand{\FeHeq}[1]{$\mbox{[Fe/H]}={#1}$}    
\newcommand{\FeHsim}[1]{$\mbox{[Fe/H]}\sim{#1}$}  
\newcommand{\FeHlt}[1]{$\mbox{[Fe/H]}<{#1}$}    
\newcommand{\FeHle}[1]{$\mbox{[Fe/H]}\le{#1}$}   
\newcommand{\ABeq}[3]{$\mbox{[#1/#2]}={#3}$}    
\newcommand{\ABlt}[3]{$\mbox{[#1/#2]}<{#3}$}    
\newcommand{\ABgt}[3]{$\mbox{[#1/#2]}>{#3}$}    
\newcommand{\ABsim}[3]{$\mbox{[#1/#2]}\sim {#3}$} 
\newcommand{\tefft}{$T_{\mbox{\scriptsize eff}}$} 
\newcommand{\teffm}{T_{\mbox{\scriptsize eff}}}  

\newcommand{\Jturnoff}{LAMOST~J125346.09$+$075343.1}
\newcommand{\Jgiant}{LAMOST~J131331.18$-$055212.4}

\newcommand{\logg}{\ensuremath{\log g}}

\newcommand{\mlp}{\ensuremath{\alpha_{\mathrm{MLT}}}}

\newcommand{\Tefft}{\ensuremath{T_{\mathrm{eff}}}}

\newcommand{\FAC}{HE~1327$-$2326} 
\newcommand{\CBB}{HE~0107$-$5240} 

\begin{document}
\SetRunningHead{Li et al.}{Ultra metal-poor stars found in LAMOST survey}

\title{High-resolution spectroscopic studies of ultra metal-poor stars found in LAMOST survey}

\author{Haining \textsc{li}\altaffilmark{1}
\thanks{This work is based on data collected at the Subaru Telescope,
which is operated by the National Astronomical Observatory of Japan.}}
\altaffiltext{1}{Key Lab of Optical Astronomy, National Astronomical Observatories,
Chinese Academy of Sciences, A20 Datun Road, Chaoyang, Beijing 100012, China}
\email{lhn@nao.cas.cn}

\author{Wako \textsc{aoki}\altaffilmark{2,3}}
\altaffiltext{2}{National Astronomical Observatory of Japan,
2-21-1 Osawa, Mitaka, Tokyo, 181-8588, Japan}
\altaffiltext{3}{Department of Astronomical Science, School of Physical Sciences,
The Graduate University of Advanced Studies (SOKENDAI), 2-21-1 Osawa, Mitaka, Tokyo 181-8588, Japan}
\email{aoki.wako@nao.ac.jp}

\author{Gang \textsc{zhao}\altaffilmark{1}}
\altaffiltext{1}{Key Lab of Optical Astronomy, National Astronomical Observatories,
Chinese Academy of Sciences, A20 Datun Road, Chaoyang, Beijing 100012, China}
\email{gzhao@nao.cas.cn}

\author{Satoshi \textsc{honda}\altaffilmark{4}}
\altaffiltext{3}{University of Hyogo, 407-2, Nishigaichi, Sayo-cho, Sayo, Hyogo, 679-5313, Japan}
\email{honda@nhao.jp}

\author{Norbert \textsc{christlieb}\altaffilmark{5}}
\altaffiltext{4}{Zentrum f{\"u}r Astronomie der Universit{\"a}t Heidelberg, Landessternwarte,
		K{\"o}nigstuhl 12, D-69117 Heidelberg, Germany}
\email{N.Christlieb@lsw.uni-heidelberg.de}

\and

\author{Takuma \textsc{suda}\altaffilmark{6}}
\altaffiltext{5}{Research Center for the Early Universe, The University of Tokyo,
Hongo 7-3-1, Bunkyo-ku, Tokyo 113-0033, Japan}
\email{suda@resceu.s.u-tokyo.ac.jp}



\maketitle

\begin{abstract}
We report on the observations of two ultra metal-poor (UMP) stars with \FeHsim{-4.0} including one new discovery.
The two stars are studied in the on-going and quite efficient project to search for
extremely metal-poor (EMP) stars with LAMOST and Subaru.
Detailed abundances or upper limits of abundances have been derived
for 15 elements from Li to Eu based on high-resolution spectra obtained with Subaru/HDS.
The abundance patterns of both UMP stars are consistent with the ``normal-population'' among
the low-metallicity stars.
Both of the two program stars show carbon-enhancement without any excess of heavy neutron-capture elements,
indicating that they belong to the subclass of CEMP-no stars,
as is the case of most UMP stars previously studied.
The \AB{Sr}{Ba} ratios of both CEMP-no UMP stars are above \ABsim{Sr}{Ba}{-0.4},
suggesting the origin of the carbon-excess is not compatible with the mass transfer from an AGB companion
where the $s-$process has operated.
Lithium abundance is measured in the newly discovered UMP star {\Jturnoff},
making it the second UMP turnoff star with Li detection.
The Li abundance of {\Jturnoff} is slightly lower than the values obtained
for less metal-poor stars with similar temperature,
and provides a unique data point at \FeHsim{-4.2} to
support the ``meltdown'' of the Li Spite-plateau at extremely low metallicity.
Comparison with the other two UMP and HMP (hyper metal-poor with \FeHlt{-5.0}) turnoff stars suggests
that the difference in lighter elements such as CNO and Na might cause notable difference
in lithium abundances among CEMP-no stars.
\end{abstract}

\section{Introduction}

The atmosphere of stars with extremely low-metallicities, i.e., extremely metal-poor
(\FeH \footnote{$[A/B]=\log(N_A/N_B)_{\star}-\log(N_A/N_B)_{\odot}$,
where $N_A$ and $N_B$ are the number densities of elements A and B respectively,
and $\star$ and $\odot$ refer to the star and the Sun respectively}
$<-3.0$, EMP) stars, preserves detailed information of chemical compositions
of the interstellar medium at the time and place that these stars were born in the early universe.
These objects, and those with even lower metallicities, e.g., ultra metal-poor (\FeHlt{-4.0}, UMP)
and hyper metal-poor (\FeHlt{-5.0}, HMP) stars are believed to provide a new observational window
to the earliest phases of evolution of the Galaxy, the formation
of the first generation of stars, and the earliest nucleosynthesis process
\citep{McWilliam1995AJ,Norris1996ApJS,Beers&Christlieb2005ARAA,Frebel&Norris2013pss}.
Detailed analysis of the chemical composition of metal-poor stars enables
us to indirectly probe the range of supernovae nucleosynthesis yields
in the early Galaxy \citep[][and references
 therein]{Heger&Woosley2010ApJ,Nomoto2013ARAA}, and to ultimately
constrain the cosmological models for the primordial nucleosynthesis as well
\citep[for details, see][]{Bromm&Yoshida2011ARAA}.

Recent studies using increasing number of metal-poor stars have made it clear that
there exist a large fraction of stars with significant enhancements of carbon,
which are usually referred to the carbon-enhanced metal-poor (CEMP) stars.
The frequency of CEMP stars is suggested to increase with decreasing metallicities
\citep{Carollo2012ApJ,Spite2013AA,Yong2013ApJb}.
Various abundance patterns of heavier elements, especially the neutron-capture elements
also indicate that there are more than one subclass of CEMP stars \citep{Beers&Christlieb2005ARAA}:
CEMP-s enriched in $s-$process elements,
CEMP-rs enriched in both s- and $r-$process elements,
and CEMP-no without enhancement in neutron-capture elements.
Detailed analysis of the elemental abundances of CEMP stars
allows us to understand the nature of their progenitor stars.
Generally, patterns in CEMP-s and CEMP-rs stars are believed to originate from
mass-transfer from a companion in the asymptotic giant branch (AGB)
with different masses and abundances.

With larger sample of CEMP stars, statistical studies have revealed that
at lower metallicities,
CEMP-no is the dominant component among all the subclasses
\citep{Norris2013ApJb,Carollo2014ApJ}.
However, the origin of CEMP-no subclass is not yet well understood
\citep[see][for a recent review]{Masseron2010AA}.
Various polluters have been suggested to explain the observed abundance pattern
of the CEMP-no stars, including mass transfer from the AGB companion \citep{Suda2004ApJ,Masseron2010AA},
faint SNe associated with the first generations of stars \citep{Umeda&Nomoto2005ApJ,Nomoto2013ARAA},
or carbon-rich winds of massive rotating EMP stars\citep{Meynet2006AA,Cescutti2013AA}, etc.

The lithium abundances provide observational constraints
to a number of important subjects such as on the origin of different CEMP subclasses,
which have been discussed in previous works \citep[e.g.,][]{Masseron2012ApJ,Spite2013AA,Hansen2014ApJ}.
The low abundance of Li in CEMP-s and CEMP-rs has been indicated to be
mainly caused by the mass transfer from the AGB companion.
However, very limited data and relevant discussions have been focused on CEMP-no stars
and their origins remain uncertain.

In the past decades, the number of metal-poor stars has been tremendously increased by
early wide-field spectroscopic surveys such as the HK survey \citep{Beers1992AJ} and Hamburg/ESO survey
\citep[HES $-$ ][]{Christlieb2008AA}, and more recently Sloan Digital Sky Survey \citep{York2000AJ}
including the Sloan Extension for Galactic Understanding and Exploration \citep[SEGUE,
][]{Yanny2009AJ,Rockosi2009astro}.
High-resolution spectroscopic follow-up observations have been performed
for the metal-poor star candidates found in these survey projects,
including the ``First Stars'' project \citep{Cayrel2004AA,Bonifacio2009AA},
the ``The Most Metal-Poor Stars'' \citep{Norris2013ApJa,Yong2013ApJa},
``Extremely Metal-Poor Stars from SDSS/SEGUE'' \citep{Aoki2013ApJ,Aoki2014arXiv},
and other projects such as \citet{Cohen2013ApJ}, \citet{Roederer2014AJ}, \citet{Jacobson2015arXiv}, etc.
These follow-up projects have provided with detailed chemical abundances of more than 300 EMP stars.
However, only 18 stars are identified to have \FeHlt{-4.0} \citep[e.g.,][]{Norris2007ApJ,Caffau2012AA,
Spite2013AA,Masseron2012ApJ,Hansen2014ApJ,Roederer2014AJ},
including the two HMP stars, \FAC ~(\FeHeq{-5.4}, \citealt{Frebel2005Nature,Aoki2006ApJ})
and \CBB ~(\FeHeq{-5.3}, \citealt{Christlieb2002Nature,Christlieb2004AA}),
and SMSS 0313$-$6708, the most iron-deficient star with an upper limit of \FeHsim{-7.1} \citep{Keller2014Nature}.
A remarkable feature of these UMP stars is the great diversity in the abundance pattern.
The number of UMP stars studied to date is, however, quite small,
and additional low-metallicity stars, especially objects with \FeHlt{-4.0},
are very important to better understand star formation, stellar evolution,
and the enrichment in the earliest phases of Galactic chemical evolution.

LAMOST \footnote{See http://www.lamost.org for more detailed information,
and the progress of the LAMOST surveys.}
\citep[the Large sky Area Multi-Object fiber Spectroscopic Telescope,
also known as Wang-Su Reflecting Schmidt Telescope or Guoshoujing Telescope;][]
{Cui2012RAA,Luo2012RAA} has started the 5-year regular survey since 2011 \citep{Zhao2012RAA}.
LAMOST combines a large aperture (4\,m), high spectrum acquiring rate
(4000 objects at one exposure), and a 5 degree field of view,
which allows us to carry out large scale spectroscopic surveys
of multiple components of the Galaxy, including metal-poor stars.
With the medium-resolution ($R = 1800$) spectroscopic data from LAMOST,
it is possible to reliably identify metal-poor stars in the survey mode,
thus evidently enhancing the searching efficiency, and provides a great opportunity
to notably enlarge the database of metal-poor stars.

We started follow-up high-resolution spectroscopy using the Subaru Telescope
for candidates EMP stars found in LAMOST.
This paper reports on detailed abundances of two UMP stars measured in the program.
The paper is organized as follows. Observations and data reduction are
addressed in Section~\ref{sec:observation};
abundance analysis is described in Section~\ref{sec:abundance},
and corresponding interpretations are described in Section~\ref{sec:results}.
The major conclusions are given in Section~\ref{sec:conclusions}.

\section{Observation and measurements}\label{sec:observation}

\subsection{Candidate selection and follow-up observation}

\begin{figure}
 \begin{center}
  \includegraphics[width=8cm]{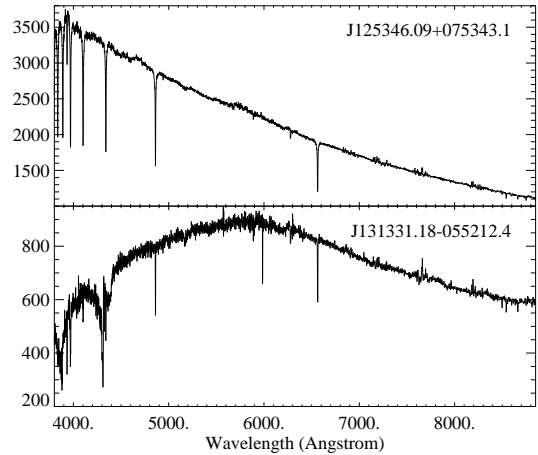}
 \end{center}
\caption{Medium-resolution spectra of the two UMP stars obtained with LAMOST.}\label{fig:LAMOST-spec}
\end{figure}

Candidates of metal-poor stars were selected from the low-resolution spectra
of the first data release of LAMOST spectroscopic survey.
The wavelength coverage ($3700$--$9100$\,{\AA}) and resolving power ($R=1800$)
of the LAMOST spectra (Fig.~\ref{fig:LAMOST-spec}) allow a robust estimation of the stellar parameters including metallicities.
Methods to determine the metallicity of an object and the selection of EMP candidates
were similar to those made by \citet{Li2015ApJ}.

Three independent methods are used to determine the metallicity based on LAMOST spectra.
The first applies an updated version of the methods described by \citet{Beers1999AJ},
which obtain {\FeH} by making use of the CaII~K line index
and the HP2 index of the H$_{\delta}$ line.
The second method matches the observed line indices to the synthetic ones.
Lick indices \footnote{http://astro.wsu.edu/worthey/html/index.table.html}
have been calculated for the observed spectra,
and compared with the synthetic line index grid to find the best match of parameters.
The third method is based on a direct comparison of the normalized observed flux
with the normalized synthetic spectra in the wavelength range $4500\,\mathrm{\AA}\le\lambda\le 5500\,\mathrm{\AA}$.
The $\chi^{2}$ minimization technique used by \citet{Lee2008AJ_SSPP1}
is adopted to find the best-matching parameters.
Any object is regarded as an EMP candidate if it is within the temperature
range $4000\,\mathrm{K}<\teffm< 7000\,\mathrm{K}$, and if at least two of
the three methods described above yield \FeHle{-2.7}.
The typical uncertainty of $\sim 0.1$ -- $0.3$\,dex
for metallicities derived from low-resolution spectra is included in the candidate selection.
All candidates which were automatically selected as described above
were visually examined to exclude false positives such as cool white dwarfs,
or spectra with reduction artifacts.


For 54 candidates of extremely metal-poor stars selected from the
LAMOST sample,``snapshot'' high-resolution spectra were acquired with
the resolving power R$=$36,000 and exposure times of 10$-$20 minutes
during the two-night run in May 2014 with Subaru/HDS as made by \citet{Aoki2013AJ}.
{\Jturnoff} and {\Jgiant} are selected from the sample
by a quick-look of the snap-shot spectra as candidates of ultra metal-poor stars.
Spectra with higher resolving power (R$=$60,000) and higher
signal-to-noise ratio covering 4000$-$6800\,{\AA} were obtained on May 10, 2014.
Exposure times for {\Jturnoff} and {\Jgiant} are both 45 minutes.
Data reduction was carried out with standard procedures
using the IRAF echelle package including bias-level correction,
scattered light subtraction, flat-fielding, extraction of spectra, and
wavelength calibration using Th arc lines. Cosmic-ray hits were removed
by the method described in \citet{Aoki2005ApJ}.

Radial velocities of the sample were obtained using the
\texttt{IRAF} procedure \texttt{fxcor}, and a synthetic spectrum
with low-metallicity was employed as a template for cross-correlation.
Equivalent widths were measured by fitting Gaussian profiles to isolated
atomic absorption lines based on the line list of \citet{Aoki2013AJ}.

\begin{table*}
\caption{Basic parameters of the 2 UMP stars and the Subaru/HDS observation.}\label{tab:obs-param}
\begin{center}
\begin{tabular}{lcccrrr}
\hline
ID                 &R.A.       &Decl.        &r   &Expt&S/N&v$_{r}$\\
                   &           &             &    &(s) &@4500\,{\AA}&(km\,s$^{-1}$)\\
\hline
{\Jturnoff}&12 53 46.09&$+$07 53 43.1&12.3&2700&165&78$\pm$0.4\\
{\Jgiant}  &13 13 31.18&$-$05 52 12.4&14.1&2700&45 &114$\pm$0.7\\
\hline
\end{tabular}
\end{center}
\end{table*}

\subsection{Stellar parameters and the ultra metal-poor stars}

Since there is not a uniform photometric system for all the LAMOST input catalogue,
we have adopted the spectroscopic method to derive the stellar parameters
of the sample. By minimizing the trend of the relationship between the derived abundances
and excitation potentials of Fe\,I lines, the effective temperatures {\tefft} of the stars
were determined. The empirical formula derived by \citet{Frebel2013ApJ}
has been adopted to correct the usually expected systematic offsets between the
spectroscopic and photometric effective temperatures.
Microturbulent velocities $\xi$ were also determined based on analysis of
Fe\,I lines, i.e., by forcing the iron abundances of individual lines to
exhibit no dependence on the reduced equivalent widths.
The sufficiently high quality of Subaru spectra allows us to detect 3 Fe\,II lines for each object.
Therefore the surface gravity {\logg} was determined by minimizing the difference
between the average abundances derived from the Fe\,I and Fe\,II lines.

Among the 54 observed candidates, there are 48 objects for which we have been able to
measure sufficient amount of atomic lines and to determine relatively reliable parameters.
The preliminary estimation of their metallicities based on the snapshot spectra
infers that there are 38 EMP stars with \FeHlt{-3.0},
including 13 with \FeHlt{-3.5}, and 2 with \FeHlt{-4.0}.
The target selection of extremely metal-poor candidates from LAMOST is thus very efficient,
with about 70\% of the observed candidates and 80\% of the targets with measured parameters
proved to be truly extremely metal-poor.
More details about the whole sample will be described in a separate paper,
and the following context focuses on the two UMP stars ({\Jturnoff} and {\Jgiant})
which have been observed with longer exposures thus achieved higher S/N ratios.

\begin{figure}
 \begin{center}
  \includegraphics[width=7.5cm]{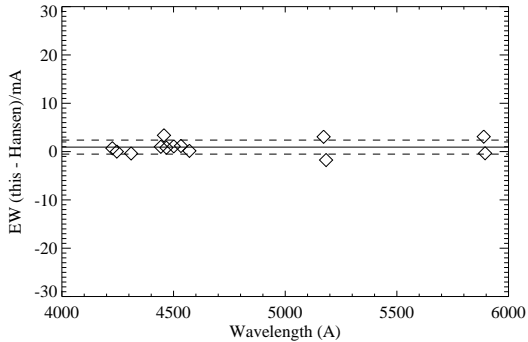}
 \end{center}
\caption{Comparison of the measured equivalent widths of common Fe I lines
by our analysis and \citet{Hansen2014ApJ}. The solid line corresponds to the
average difference of the measured equivalent width,
and the dashed lines refer to the standard deviation
of the difference, i.e., $0.9\pm 1.4$\,m$\AA$.}\label{fig:J1313_EW_compare}
\end{figure}

{\Jgiant} has been independently discovered to be an EMP star, HE\,1310$-$0520
by HES and followed-up with UVES.
After our follow-up observation with Subaru, results of abundance measurements for this objects
together with three other UMP stars was published \citep{Hansen2014ApJ}.
Whereas only 18 Fe\,I lines were used by \citet{Hansen2014ApJ} for parameter determination,
we used more than 40 Fe\,I lines for our estimation.
For the common Fe\,I  lines, Fig.~\ref{fig:J1313_EW_compare}
compares the equivalent widths that have been adopted by us and \citet{Hansen2014ApJ}.
A good agreement is found between the two measurements, with a mean difference of $-$0.9\,m{\AA} with $\sigma=1.4$\,m{\AA}.

\begin{table*}
\caption{Basic parameters of the 2 UMP stars and the Subaru/HDS observation.}\label{tab:stellar-param}
\begin{center}
\begin{tabular}{lcccccccc}
\hline
ID                 &\multicolumn{4}{c}{Subaru Measurement}&&\multicolumn{3}{c}{LAMOST measurement}\\
\cline{2-5}\cline{7-9}
                   &{\tefft}&{\logg}&{\FeH}&$\xi$&&{\tefft}&{\logg}&{\FeH}\\
                   &(K)&    &     &(km\,s$^{-1}$)&&(K)&      &\\
\hline
{\Jturnoff}&6030$\pm$135&3.65$\pm$0.16&$-$4.02$\pm$0.06&1.4$\pm$0.12&&6034&2.75&$-$3.97\\
{\Jgiant}&4750$\pm$94 &1.60$\pm$0.21&$-$4.12$\pm$0.13&1.5$\pm$0.07&&4630&1.32&$-$4.23\\
\hline
\end{tabular}
\end{center}
Note: The error associated with the stellar parameters from Subaru measurement
corresponds to the internal uncertainties derived from the spectroscopic analysis.
\end{table*}

Table~\ref{tab:stellar-param} presents the stellar parameters of {\Jturnoff} and {\Jgiant}
which are derived from Subaru and LAMOST spectra.
For {\Jgiant}, the derived metallicities are consistent with the value from \citet{Hansen2014ApJ},
but there is a notable difference of $-$250\,K in {\tefft}.
To further investigate the cause of the difference, we have also checked the photometric temperature
of the two objects. The $V-K$ color, as well as $J-K$, is available for {\Jgiant},
while only $J-K$ is available for {\Jturnoff}. Considering the calibration by \citet{Alonso1999AAS,Alonso2001AA},
the derived temperatures are 6035\,K for {\Jturnoff} (based on $J-K$) and 4880\,K for {\Jgiant} (based on $V-K$).
Hence, for both of the two UMP stars, the spectroscopic temperature well agrees
with the photometric value as well as that derived from LAMOST spectra.
However, for {\Jgiant}, the spectroscopic temperature is notably lower than
that derived by \citet{Hansen2014ApJ} through
fitting the spectrophotometric observations with model atmosphere fluxes.
We hence suspect that the difference may come from the different method of stellar parameter determination,
which tends to be larger for cooler giants, e.g., for objects with strong molecular absorption
such as the case of {\Jgiant} (as can be seen from Fig.~\ref{fig:LAMOST-spec}).

\section{Abundance determination}\label{sec:abundance}

The 1D plane-parallel, hydrostatic model atmospheres of the ATLAS NEWODF grid of \citet{Castelli&Kurucz2003IAUS}
are used for the abundance analysis, assuming a mixing-length parameter of $\mlp=1.25$, no convective
overshooting, and local thermodynamic equilibrium. We use an updated
version of the abundance analysis code MOOG \citep{Sneden1973ApJ}, which
does not treat continuous scattering as true absorption,
but as a source function which sums both absorption and scattering \citep{Sobeck2011AJ}.
The photospheric Solar abundances of \citet{Asplund2009ARAA} are adopted when
calculating [X/H] and [X/Fe] abundance ratios.

\subsection{Methods}

The lithium and carbon abundances of the program stars were respectively derived
by matching the observed Li\,I 6707.8\,{\AA} doublet and CH A-X band head at 4310\,{\AA}
~(i.e., the G-band) to the synthetic spectra.

For 14 other elements including Na, Mg, Si, Ca, Sc, Ti, Cr, Fe, Co, Ni, Zn, Sr, Ba, and Eu
abundances (or upper limits) were computed using the measured equivalent widths
(or estimated EW errors) with the derived stellar parameters.
If the elemental abundance was derived from a single line,
or if the derived abundances deviated by more than 3$\,\sigma$ from the average
values computed for an atomic species from multiple lines,
the abundances were checked with spectral synthesis, and modified if necessary.
For most of the used lines, synthetic spectra with abundances
derived from the measured equivalent widths match the observed spectra well.
However, for lines that suffer from blending or problems in setting the continuum level,
the observed spectral line was not properly reproduced with the abundance derived from equivalent widths.
In such cases, the abundance derived from spectral synthesis was adopted.
For lines with no detectable features but with reasonable S/N,
a 3$\sigma$ upper limit of corresponding elemental abundace
was estimated based on the \citet{Bohlin1983ApJS} formular,
i.e., $\sigma=wn_{pix}^{1/2}/(\mbox{S/N})$,
where $w$ is the pixel width, and $n_{pix}$ corresponds to number of pixels across the line.
Note that no useful upper limits could be derived for Si of {\Jturnoff}
due to difficulty in defining the continuum nearby Si 4102\,{\AA}.
The derived abundances (upper-limits) of the program stars are listed in Table~\ref{tab:abundance},
which also includes the number of lines that have been used for deriving the abundance, $N$,
together with the abundance error as described in the following text.

\begin{table*}
\caption{Elemental abundances of the sample stars.}\label{tab:abundance}
\begin{center}
\begin{tabular}{lrrrrcrrrrrr}
\hline\hline
   &\multicolumn{4}{c}{{\Jturnoff}}&&\multicolumn{4}{c}{{\Jgiant}}&&Sun\\
\cline{2-5}\cline{7-10}\cline{12-12}
Ion&log\,$\epsilon$(X)&$\sigma$&\AB{X}{Fe}&$N$&&log\,$\epsilon$(X)&$\sigma$&\AB{X}{Fe}&$N$&&log\,$\epsilon$(X)\\
\hline
Li    &    1.80&0.14&     ...& 1&& $<$0.68& ...&    ... & 1&&   1.05\\
C     &    6.00&0.19&    1.59& 1&&    6.15&0.46&    1.83& 1&&   8.43\\
Na    &    2.02&0.18& $-$0.20& 2&&    2.07&0.14& $-$0.06& 2&&   6.24\\
Mg    &    3.82&0.10&    0.24& 2&&    3.83&0.16&    0.34& 3&&   7.60\\
Si    &    ... & ...&    ... & 0&&    3.85&0.16&    0.45& 1&&   7.51\\
Ca    &    2.69&0.18&    0.37& 2&&    2.37&0.20&    0.14& 2&&   6.34\\
Sc    & $-$0.66&0.12&    0.21& 1&& $-$1.13&0.19& $-$0.17& 2&&   3.15\\
Ti    &    1.43&0.12&    0.50& 9&&    1.20&0.16&    0.36& 9&&   4.95\\
Cr    &    1.56&0.15& $-$0.06& 4&&    1.05&0.20& $-$0.48& 2&&   5.64\\
Fe\,I &    3.48&0.13&    ... &18&&    3.39&0.22&    ... &41&&   7.50\\
Fe\,II&    3.47&0.10&    ... & 3&&    3.38&0.16&    ... & 3&&   7.50\\
Co    & $<$1.75& ...& $<$0.78& 1&&    0.91&0.22&    0.03& 1&&   4.99\\
Ni    & $<$2.87& ...& $<$0.67& 0&&    1.91&0.18& $-$0.20& 1&&   6.22\\
Zn    & $<$2.06& ...& $<$1.52& 1&& $<$1.71& ...& $<$1.26& 1&&   4.56\\
Sr    & $-$0.96&0.11&    0.19& 2&& $-$2.05&0.20& $-$0.81& 2&&   2.87\\
Ba    &$<-$2.14& ...&$<-$0.30& 1&& $-$2.50&0.21& $-$0.57& 2&&   2.18\\
Eu    &$<-$1.90& ...&$<$ 1.60& 1&&$<-$3.00& ...&$<$ 0.59& 1&&   0.52\\
\hline
\end{tabular}
\end{center}
Note: $\sigma$ refers to the total uncertainty of the abundance errors,
as described in the text.
\end{table*}

\begin{figure}
 \begin{center}
  \includegraphics[width=7.5cm]{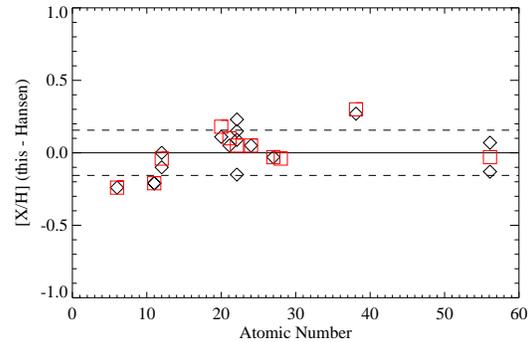}
 \end{center}
\caption{Comparison of abundances (\AB{X}{H}) of elements from C to Ba
derived for {\Jgiant} (HE~1310$-$0520) by our analysis and that of \citet{Hansen2014ApJ}.
The diamonds represent difference of abundance for individual lines, and the squares refer to the averaged
abundances for each specie. The dashed lines refer to the standard deviation of the abundance
difference ($\pm 0.16$\,dex).}\label{fig:J1313_abun_compare}
\end{figure}

We have also compared the abundances \AB{X}{H} of {\Jgiant} with those from \citet{Hansen2014ApJ},
as shown in Fig.~\ref{fig:J1313_abun_compare}. The average difference between the abundance
from this work and that from \citet{Hansen2014ApJ} is 0.01$\pm$0.16,
indicating a very good agreement between the results of the two studies.
We note that the {\Tefft} adopted in the present work is about 250\,K lower than
that of \citet{Hansen2014ApJ}. This would result in lower Fe abundances based on Fe~I lines (see below).
However, the smaller microturbulent velocity adopted in our analysis mostly compensate the abundance difference.
The difference of $\xi$ may be caused by using different Fe~I lines in the two sets of analysis.
We note that Fe~I lines used in our study is larger in number than those of \citet{Hansen2014ApJ},
and also cover a wider range of equivalent widths.

\subsection{Uncertainties}

The error of the derived abundances of the sample mainly comes from two aspects,
which includes the uncertainties of the equivalent width measurements,
and those caused by the uncertainties of stellar parameters.
In the case of equivalent width measurement,
the dispersion around the average abundance was used to present random error
in the measurements of the equivalent widths.
The errors relevant to the uncertainties of the equivalent width measurement
are listed in the columns of $\Delta$EW in Table~\ref{tab:abun-error}.
The abundance uncertainties associated with the uncertainties of
the stellar parameters were estimated by individually varying {\tefft} by $+$150\,K,
{\logg} by $+$0.3\,dex, {\FeH} by $+$0.3\,dex, and $\xi$ by $+$0.3\,km\,s$^{-1}$
in the stellar atmospheric model.
The adopted uncertainties of parameters are conservative estimations
of the internal uncertainties derived from the spectroscopic analysis as shown in Table~\ref{tab:stellar-param},
and are comparable with previous experience using high-resolution (R$\sim$30,000)
spectra and spectroscopic method to determine stellar parameters of EMP stars \citep[e.g.][]{Placco2014ApJa}.
In Table~\ref{tab:abun-error},
columns $\Delta${\tefft}, $\Delta${\logg}, $\Delta${\FeH}, and $\Delta \xi$
summarize corresponding quantities in abundance uncertainties.
The total uncertainty of the errors, which was computed
as the quadratic sum of the above mentioned aspects,
as shown in the column of $\sigma$ in Table~\ref{tab:abundance},
and that of ``Total'' in Table~\ref{tab:abun-error}.

\begin{table*}
\caption{Abundance errors of the sample star. Details are described in the text.}\label{tab:abun-error}
\begin{center}
\begin{tabular}{rrrrrrrcrrrrrr}
\hline\hline
   &\multicolumn{6}{c}{{\Jturnoff}}&&\multicolumn{6}{c}{{\Jgiant}}\\
\cline{2-7}\cline{9-14}
Ion&$\Delta$EW&$\Delta$\,{\tefft}&$\Delta$\,{\logg}&$\Delta$\,{\FeH}&$\Delta$\,$\xi$&Total&
&$\Delta$EW&$\Delta$\,{\tefft}&$\Delta$\,{\logg}&$\Delta$\,{\FeH}&$\Delta$\,$\xi$&Total\\
\hline
Li   &0.10&0.10&   0.00&   0.00&   0.00&0.14& & ...& ...&    ...& ...&    ...& ...\\
C    &0.15&0.10&   0.00&   0.05&   0.05&0.19& &0.20&0.40&$-$0.10&0.05&   0.00&0.46\\
Na   &0.14&0.11&   0.00&   0.00&   0.00&0.18& &0.00&0.14&$-$0.02&0.00&$-$0.03&0.14\\
Mg   &0.01&0.10&   0.00&   0.00&$-$0.03&0.10& &0.03&0.13&$-$0.03&0.00&$-$0.09&0.16\\
Si   & ...& ...&    ...&    ...&    ...& ...& &0.05&0.15&   0.00&0.00&   0.00&0.16\\
Ca   &0.14&0.10&$-$0.01&$-$0.01&$-$0.06&0.18& &0.03&0.15&$-$0.04&0.00&$-$0.12&0.20\\
Sc   &0.05&0.09&   0.06&   0.00&   0.00&0.12& &0.07&0.11&   0.07&0.00&$-$0.12&0.19\\
Ti   &0.09&0.06&   0.06&   0.00&   0.00&0.12& &0.10&0.08&   0.08&0.00&$-$0.04&0.16\\
Cr   &0.05&0.14&   0.00&   0.00&$-$0.01&0.15& &0.00&0.19&$-$0.03&0.00&$-$0.04&0.20\\
FeI  &0.03&0.13&   0.00&   0.00&$-$0.02&0.13& &0.13&0.18&$-$0.03&0.00&$-$0.01&0.22\\
FeII &0.08&0.02&   0.06&   0.00&   0.00&0.10& &0.12&0.03&   0.10&0.00&$-$0.01&0.16\\
Co   & ...& ...&    ...&    ...&    ...& ...& &0.08&0.20&$-$0.02&0.00&$-$0.01&0.22\\
Ni   & ...& ...&    ...&    ...&    ...& ...& &0.08&0.16&$-$0.02&0.00&$-$0.01&0.18\\
Sr   &0.01&0.09&   0.06&   0.00&$-$0.03&0.11& &0.06&0.12&   0.07&0.00&$-$0.13&0.20\\
Ba   & ...& ...&    ...&    ...&    ...& ...& &0.14&0.13&   0.07&0.00&$-$0.03&0.21\\
\hline
\end{tabular}
\end{center}
\end{table*}

\section{Restuls and interpretations}\label{sec:results}

\subsection{Abundance trends from C through Zn}

\begin{figure*}
 \begin{center}
  \includegraphics[width=12cm]{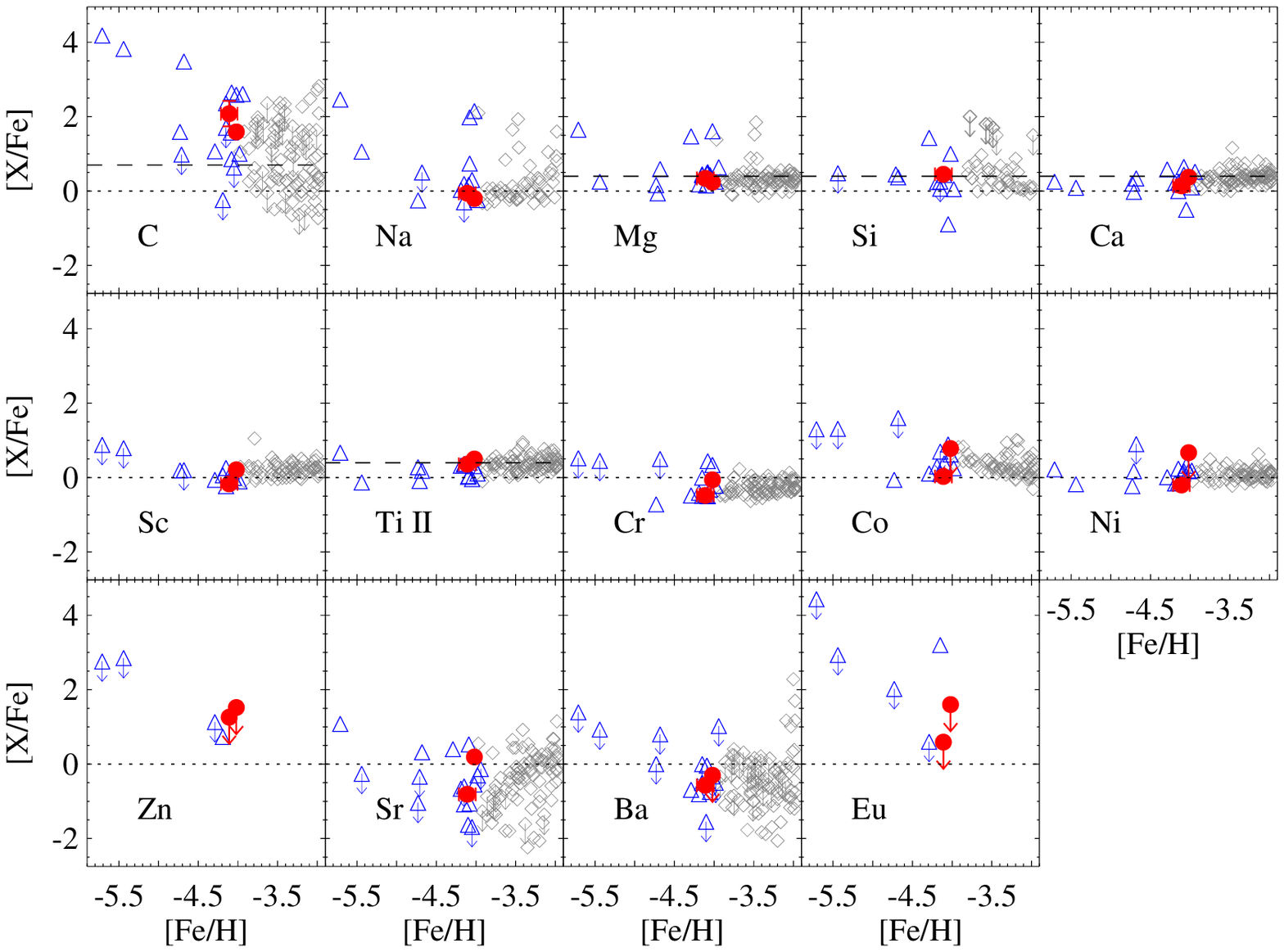}
 \end{center}
\caption{\AB{X}{Fe} vs. {\FeH} of 14 elements from C to Eu. For carbon, the dashed line
refers to \ABsim{C}{Fe}{+0.7}, i.e., a conservative division of carbon-enhanced and carbon-normal stars
\citep[e.g., ][]{Aoki2007ApJ}. For all $\alpha$ elements (Mg, Si, Ca, and Ti),
the canonical value of \ABsim{$\alpha$}{Fe}{+0.4} for the halo stars \citep{Mcwilliam1997ARAA}
is plotted for reference. Diamonds refer to extremely metal-poor stars from \citet{Yong2013ApJa};
triangles refer to other ultra and hyper metal-poor stars from literatures;
filled circles are the two program stars.
Arrows correspond to upper or lower limits.}\label{fig:abun_C2Ba}
\end{figure*}

For 11 elements from C to Zn, abundance ratios of \AB{X}{Fe} of the program stars together with
those of other known 15 UMP and 2 HMP stars from literatures are plotted against {\FeH} in Fig.~\ref{fig:abun_C2Ba}.
Abundances of the extremely metal-poor stars from \citet{Yong2013ApJa} are also included for comparison.
Note that the most iron-poor star SMSS\,0313$-$6708 with only an upper limit of {\FeH} of $-$7.1
is not included in the discussion considering the scale of the plots.

Both of the two program stars are carbon-enhanced,
as for 14 out of the 18 collected sample of UMP and HMP stars.
This provides further support for the picture of high frequency of carbon enhancement
in the early universe pointed out by previous studies \citep[e.g.,][]{Norris2013ApJb,Placco2014ApJb}.
Such dominant enrichment of carbon at the beginning of the chemical evolution
can be explained by the scenario, e.g., for low-mass star-formation proposed by \citet{Frebel2007MNRAS},
which indicates a cooling process through fine-structure lines of carbon and oxygen.

When investigating the carbon abundances of unevolved dwarf and turnoff metal-poor stars,
\citet{Spite2013AA} has found a possible ``plateau'' or an upper limit of $A\mbox{(C)}=6.8$
for carbon-rich metal-poor stars with \FeHlt{-3.4}.
For the unmixed turnoff star in our sample, {\Jturnoff}, we have derived a carbon abundance
of $A\mbox{(C)}=6.0$, which is well below the ``plateau'' value,
while \citet{Hansen2014ApJ} has derived $A\mbox{(C)}=7.2$ for the turnoff UMP HE~0233$-$0343 in their sample.
Therefore additional data of less evolved CEMP stars are required
before one can conclude whether there exists such a carbon ``plateau'' at low metallicities.

Aside from the enhancement of carbon, the two program stars in our work show no excess or deficiency of
$\alpha$-elements or iron-peak elements, but follow the general trend of abundance ratios
in other elements as found in ``normal'' EMP stars.
This is also consistent with the indication by larger samples of EMP stars
\citep[e.g., ][Li et al. 2015]{Yong2013ApJa}, that there is a ``normal'' population
dominating the elemental abundance pattern even at the extremely low-metallicity region,
except for the carbon-excesses found in many EMP stars.

The abundance ratios of the two stars are very similar in general.
Some notable differences are found for Sc and Cr.
The \AB{Sc}{Fe} of the red-giant {\Jgiant} is about $-$0.3\,dex
lower than that of the turn-off star {\Jturnoff},
which is consistent with the discrepancy found by \citet{Bonifacio2009AA}
when comparing EMP turnoff and giant stars.
However, they also indicate that the abundance difference cannot be explained by the granulation effect.
A low \AB{Cr}{Fe} is also obtained for {\Jgiant},
whereas \ABsim{Cr}{Fe}{0} is derived for {\Jturnoff}.
Such a discrepancy between giants and dwarf stars have been suggested by
previous studies \citep{Lai2008ApJ,Bonifacio2009AA}
that have suggested the non-LTE effect in neutral Cr lines.

\subsection{Neutron-capture elements}

The wavelength coverage and quality of the spectra enable us to
measure abundances of Sr and Ba, which represent light and heavy
neutron-capture elements. The abundances of Sr are derived for both
objects, while the Ba abundance is determined for the giant {\Jgiant}.
Only an upper limit of the Ba abundance is estimated for the turn-off
star {\Jturnoff}.  As can be seen from the last two panels of
Fig.~\ref{fig:abun_C2Ba}, the number of UMP stars with detected Sr and
Ba is still very small.  The abundance of these elements in UMP and
HMP stars are very important to understand the neutron-capture
processes in the early universe and the origin of carbon-excess for CEMP stars.

The abundance ratios of \AB{Sr}{Fe} and \AB{Ba}{Fe} of the two program
stars are all around or slightly below the Solar value.
Such low abundances of neutron-capture elements indicate no significant
contribution of $s-$process to these stars, and that these stars belong
to the CEMP-no sub-class
\citep{Beers&Christlieb2005ARAA,Sivarani2006AA}. This result supports
the findings of previous studies that the significant fraction of CEMP
stars are classified into CEMP-no at extremely low metallicity
\citep{Aoki2007ApJ,Norris2013ApJb}.

\begin{figure*}
 \begin{center}
  \includegraphics[width=10cm]{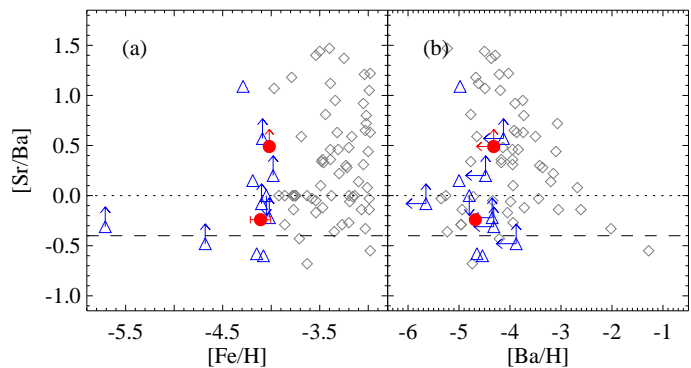}
 \end{center}
\caption{\AB{Sr}{Ba} vs. {\FeH} (a) and \AB{Ba}{H} (b) of the program stars.
Meanings of different symbols are the same as in Fig.~\ref{fig:abun_C2Ba}.
The dashed line corresponds to \ABsim{Sr}{Ba}{-0.4}, i.e., the upper limit
in case of mass transfer from an AGB companion \citep{Spite2013AA}.}\label{fig:abun_SrBa}
\end{figure*}

The abundance ratios of the two elements, \AB{Sr}{Ba}, are useful to
constrain the possible nucleosynthesis processes that have taken place
in the progenitors of these stars \citep{Francois2007AA,Aoki2013ApJ}.
Since production of heavy neutron-capture elements (e.g., Ba)
relative to light ones (e.g., Sr) is efficient in the main s-porcess
at low metallicity due to high ratios of neutron to seed nuclei
\citep[e.g.][]{Busso1999ARAA},
a low \AB{Sr}{Ba} ratio, e.g., \ABlt{Sr}{Ba}{-0.4}, is expected,
as shown by the dashed line in Fig.~\ref{fig:abun_SrBa} \citep{Spite2013AA}.
The \AB{Sr}{Ba} of the two objects higher than this limit (Fig.~\ref{fig:abun_SrBa})
excludes the main $s-$process from possible origins of neutron-capture elements.
Besides, as shown in Fig.~\ref{fig:abun_SrBa}b, the \AB{Sr}{Ba} of the two stars
are within the range found by previous studies for stars with low \AB{Ba}{H}.

Previous works have observed the absence of stars with \ABgt{Sr}{Ba}{0}
at approximately \FeHsim{-3.6} or lower, as can be seen in Fig.~\ref{fig:abun_SrBa}a.
The high Sr abundances found
in many EMP stars with $-3.6\lesssim$ \FeH $\lesssim -3$ are
interpreted as significant contributions of neutron-capture process
that only yields light neutron-capture elements, e.g., weak $r-$process
\citep{Wanajo2006NuPhA}; Lighter Element Primary Process \citep[LEPP,][]{Travaglio2004ApJ}.
The cut-off suggests that such a process is not efficient at the
lowest metallicity range, e.g. \FeHlt{-3.6}. The relatively high
lower limit of \AB{Sr}{Ba} of {\Jturnoff} suggests, however, that the
cut-off is located at lower metallicity (\FeHlt{-4}) if it exists at all.

We note, however, that {\Jturnoff} is a CEMP-no star
whose metallicity may not be well represented simply by the Fe abundance,
but the excess of carbon should be taken into consideration
in the comparisons with most of other EMP stars with \FeHlt{-3.6} shown in Fig.~\ref{fig:abun_SrBa}.
The excess of Sr in this CEMP-no star also recalls the
detection of Sr in the HMP star HE~1327$-$2326
(\ABeq{Sr}{Fe}{+1}). Although the lower limit of \AB{Sr}{Ba} of this
object is still low due to the high upper-limit of the Ba abundance,
this object could also have high Sr/Ba ratios.
However, the enhancement of Sr among CEMP-no stars are not especially associated with binaries,
e.g., close binaries that favor mass transfer can be excluded in most cases \citep{Starkenburg2014MNRAS},
while the connection with the carbon-excess might be able to probe the process to
produce light neutron-capture elements at low metallicity.

\subsection{Lithium}

\begin{figure}
 \begin{center}
  \includegraphics[width=8cm]{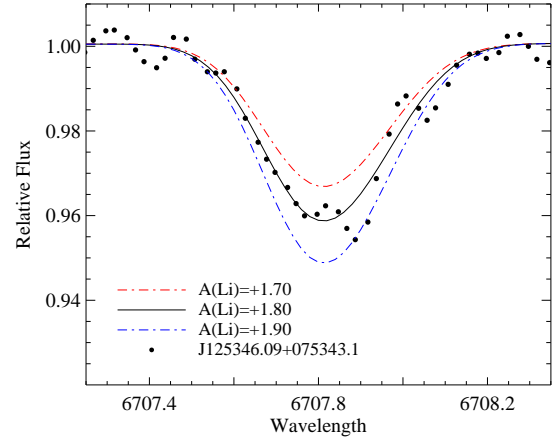}
 \end{center}
\caption{Subaru spectrum of {\Jturnoff} in the region of Li 6707\,{\AA}.
The dots represent the observed spectra;
the solid line refers to the best fit with $A\mbox{(Li)=1.80}$,
and the upper and lower dashed-dotted lines correspond to synthetic spectra
with changes of 1$\sigma$-fitting-uncertainty in $A\mbox{(Li)}$.}\label{fig:synfit_Li}
\end{figure}

Lithium is only detected in the warmer one of the program stars
{\Jturnoff}, with $A\mbox{(Li)}=1.80$\footnote{$A\mbox{(Li)}$ is
  defined as $A\mbox{(Li)}=\log(N\mbox{(Li)}/N\mbox{(H)})+12$.}
(as shown in Fig.~\ref{fig:synfit_Li}).
This is the second lowest-metallicity turnoff star with Li detection,
following HE~0233$-$0343 with \FeHsim{-4.7} and
$A\mbox{(Li)=1.77}$ \citep{Hansen2014ApJ}. A quite low upper limit,
$A\mbox{(Li)} < 0.68$, is derived for the other object
{\Jgiant}. The derived upper limit is close to that from
\citet{Hansen2014ApJ}.  Such low value is expected for red giants
like {\Jgiant}, because they have already undergone the first dredge
up and the surface material is mixed with internal one with
depleted Li.

For further discussion on lithium in EMP/UMP stars, we limit the
{\tefft} range to exclude severely Li-depleted stars such as the
evolved giant {\Jgiant}.  Among a variety of criteria of {\tefft}
adopted in previous studies
\citep[e.g.][]{Spite1982AA,Shi2007AA,Melendez2010AA} we adopt $\teffm
> 5800$\,K. This range includes the UMP turn-off star SDSS~1029+1729
with a low upper limit of the Li abundance \citep{Caffau2012AA}.

\begin{figure*}
 \begin{center}
  \includegraphics[width=14cm]{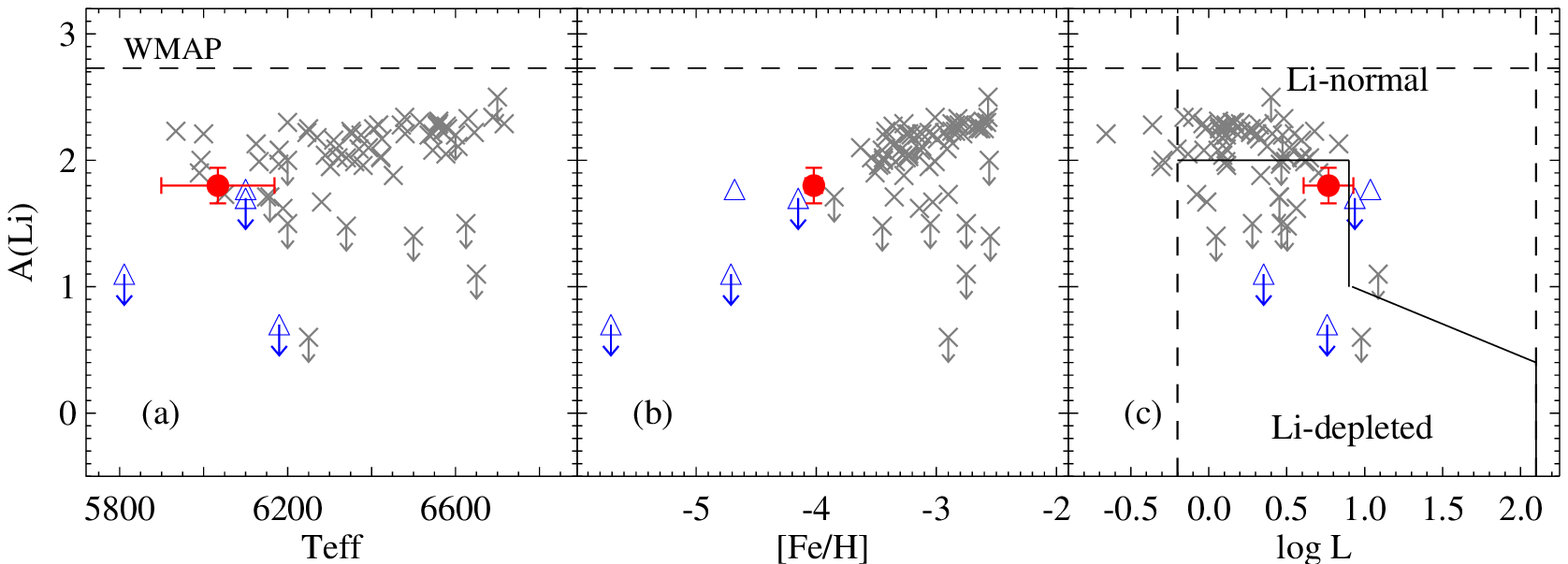}
 \end{center}
\caption{The abundance of Li, $A\mbox{(Li)}$ vs. {\tefft} (a), {\FeH} (b), and $\log$\,L (c).
Crosses are metal-poor turnoff/dwarf stars from literatures \citep{Aoki2009ApJ,Sbordone2010AA,Bonifacio2012AA,Masseron2012ApJ}.
The horizontal dashed line in all panels indicates the estimated primordial Li abundance
based on WMAP results \citep{Cyburt2008JCAP}.
The solid line in (c) corresponds to the empirical limit between Li-normal and Li-depleted
stars as defined by \citet{Masseron2012ApJ}; while the two dashed lines embrace
the space which has been considered for discussions in \citet{Masseron2012ApJ}.}\label{fig:abun_Li}
\end{figure*}

Li abundances of EMP/UMP stars are shown in Fig.~\ref{fig:abun_Li} as a function of
{\tefft}, {\FeH} and luminosity.
Metal-poor turnoff stars and dwarfs from literatures
\citep{Aoki2009ApJ,Sbordone2010AA,Bonifacio2012AA,Masseron2012ApJ} are
also included for comparison in Fig.~\ref{fig:abun_Li}. To focus on
lower metallicities, only objects with \FeHlt{-2.5} are shown in the
plot.

In Fig.~\ref{fig:abun_Li}a, the observed plateau of lithium abundance
($A$(Li)$\sim 2.2$) is clearly seen, which is known to be about
0.5\,dex lower than the model prediction by the big-bang nucleosynthesis
based on the cosmological parameters determined from WMAP measurement,
whereas a fraction of stars show lower values.
The UMP turnoff star {\Jturnoff} in our sample is
slightly below the observed plateau, compared with objects with
similar temperatures but with higher metallicities.

The Li abundance of {\Jturnoff} is similar to that of the UMP
turn-off star HE~0233$-$0343 with \FeHeq{-4.7} \citep{Hansen2014ApJ}.
Fig.~\ref{fig:abun_Li}b shows the Li abundances as a function of Fe abundances.
Our measured lithium abundance of {\Jturnoff} privides with a unique data
point in the metallicity range around \FeH$\sim -4.2$, suggesting
decreases of Li abundances with decreasing metallicitiy, which is
sometimes called the Spite plateau ``meltdown'' below
\FeHsim{-2.8} \citep{Sbordone2012MSAIS}. However,
the Li detection for only two objects below \FeHsim{-4.0} does
not allow us to investigate in more details about the explanation for such meltdown.

It should be noted that {\Jturnoff} and HE~0233$-0343$ are both
CEMP-no stars. Li abundances of CEMP stars were studied by
\citet{Masseron2012ApJ} for an extended sample of more than 40
stars. Most of the stars in their sample are CEMP-s or CEMP-rs stars,
which would be affected by the mass transfer from AGB companions.
\citet{Masseron2012ApJ} showed that CEMP-no stars in their sample are
Li-depleted, but the sample size is still small to derive any conclusions.

\begin{figure}[h]
 \begin{center}
  \includegraphics[width=8cm]{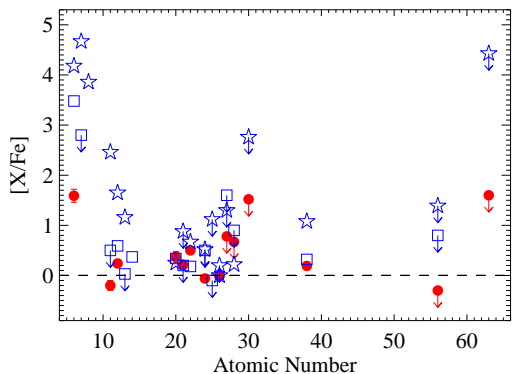}
 \end{center}
\caption{Comparison of the abundance pattern among {\Jturnoff} (filled circles), HE~0233$-$0343 (squares),
and HE~1327$-$2326 (pentagrams). The dashed line indicates the solar abundance.}\label{fig:abun_pattern}
\end{figure}

It is noted that in all plots of Fig~\ref{fig:abun_Li}, except the
difference in the metallicity, {\Jturnoff} is quite similar to
HE~0233$-$0343, including the lithium abundances, evolutionary status
(luminosity), and the fact of both being CEMP-no
\citep{Hansen2014ApJ}. These objects are also similar to the HMP
CEMP-no turnoff star HE~1327$-$2326. However, no Li is detected in
this HMP star, and a tight upper limit of $A$(Li)$ < 0.7$\,dex is derived
\citep{Frebel2008ApJ}.  Elemental abundances of the three objects are
compared in Fig.~\ref{fig:abun_pattern}. Although these three objects
are carbon-enhanced stars with no clear excess of the heavy
neutron-capture element Ba, the abundance ratios of carbon, Na, and
$\alpha$-elements are quantitatively different between the two UMP
stars and HE~1327$-$2326, which might be related to the large
difference of the Li abundance of HE~1327$-$2326 from the other two stars.

\section{Conclusions}\label{sec:conclusions}

We present high-resolution and high-quality spectroscopic observations of two UMP stars
with \FeHlt{-4.0}, adding one new discovery to the other dozen of such rare objects.
The two objects have been selected from the LAMOST spectroscopic survey,
and followed-up using Subaru/HDS. These are part of a larger project on searching for
EMP stars with LAMOST and Subaru, and the current result proves the project to be quite efficient,
reaching an 80\% success in identifying EMP stars in the Galaxy.

Detailed abundance analyses (or upper limit estimation) are performed
for 15 elements including Li, C, Na, four $\alpha-$elements (Mg, Si, Ca, and Ti),
five iron-peak elements (Sc, Cr, Co, Ni, and Zn),
and three neutron-capture elements (Sr, Ba, and Eu).
Both program stars are carbon-enhanced, in consistency with previous studies
which suggest higher fraction of CEMP stars in lower metallicity range.
Moreover, neither of the two objects are enhanced in neutron-capture element,
supporting previous observational results
that the subclass of CEMP-no dominates the extremely low-metallicity region.
The origins of the enhancement of Sr in these objects cannot solely be explained
by the mass transfer from an AGB companion, while the connection with their carbon excesses
may shed light on productions of light neutron-capture elements at early times.

The lithium abundance is determined for the turnoff star {\Jturnoff},
which makes it the second UMP turnoff star with Li detection.
This newly discovered UMP turnoff star is located at a unique position around \FeHsim{-4.2}
to support the so-called ``meltdown'' of the Li plateau at extremely low metallicities.
It is noted that all known UMP and HMP turnoff stars are CEMP-no,
except for SDSS 1029+1729 \citep{Caffau2012AA}, and depleted in lithium,
which is consistent with the suggestion by \citet{Masseron2012ApJ}.
The origin of such trend is not yet clear, and larger sample of
unevolved CEMP-no stars with Li detections would be important.
When comparing with other UMP and HMP turnoff/dwarf stars,
the turnoff star of our sample, {\Jturnoff} shows very close Li abundance
to that of HE~0233$-$0343 ($A\mbox{(Li)}\sim1.80$),
while the HMP turnoff star HE~1327$-$2326 has no Li detection
but a quite low upper limit of $A\mbox{(Li)}<0.70$.
By comparing their abundance patterns, it is shown that HE~1327$-$2326
is extremely enhanced in Na, Mg, and Sr, in contrast to the ``normal'' pattern
of {\Jturnoff} and HE~0233$-$0343. Different polluters at their birth time and places
may have been causing the observed discrepancies in Li abundances.

The huge amount of data which will be obtained through LAMOST spectroscopic survey,
and joint collaborations with Subaru will enable us to significantly
enlarge the sample of UMP stars, so as to explore the nature of the nucleosynthesis and
chemical enrichment at the very beginning of the Universe.

\begin{ack}
We are grateful to the anonymous referee who made valuable suggestions and help improve the paper.
H.N.L. and G.Z. acknowledge supports by NSFC grants No. 11103030, 11233004, and 11390371.
W.A. and T.S. are supported by the JSPS Grant-in-Aid for Scientific Research (S:23224004).
S.H. is supported by the JSPS Grant-in-Aid for Scientific Research (c:26400231).
N.C. acknowledges support from Sonderforschungsbereich 881
``The Milky Way System'' (subproject A4) of the German Research Foundation (DFG).
Guoshoujing Telescope (the Large Sky Area Multi-Object Fiber Spectroscopic Telescope, LAMOST)
is a National Major Scientific Project built by the Chinese Academy of Sciences.
Funding for the project has been provided by the National Development and Reform Commission.
LAMOST is operated and managed by the National Astronomical Observatories, Chinese Academy of Sciences.
\end{ack}

\bibliographystyle{apj} 
\bibliography{mpstar,lamost} 

\end{document}